# Chitosan/alginate bionanocomposites adorned with mesoporous silica nanoparticles for bone tissue engineering

Satar Yousefiasl [1], Hamed Manoochehri [2], Pooyan Makvandi [3], Saeid Afshar [2], Erfan Salahinejad [4], Pegah Khosraviyan [5], Massoud Saidijam [2], Sara Soleimani Asl [6], Esmaeel Sharifi [2,7]

[1] School of Dentistry, Hamadan University of Medical Sciences, Hamadan 6517838736, Iran

[2] Research Center for Molecular Medicine, Hamadan University of Medical Sciences, Hamadan, Iran

[3] Centre for Materials Interface, Istituto Italiano Di Tecnologia, viale Rinaldo Piaggio 34, 56025 Pontedera, Pisa, Italy

[4] Faculty of Materials Science and Engineering, K. N. Toosi University of Technology, Tehran, Iran

[5] Medical Plants Research Center, Basic Health Sciences Institute, Shahrekord University of Medical Sciences, Shahrekord, Iran

[6] Department of Anatomy, School of Medicine, Hamadan University of Medical Sciences, Hamadan, Iran

[7] Department of Tissue Engineering and Biomaterials, School of Advanced Medical Sciences and Technologies, Hamadan University of Medical Sciences, Hamadan, Iran

* Esmaeel Sharifi: esmaeel.sharifi@gmail.com

## Abstract

The regeneration of oral and craniofacial bone defects ranging from minor periodontal and peri-implant defects to large and critical lesions imposes a substantial global health burden. Conventional therapies are associated with several limitations, highlighting the development of a unique treatment strategy, such as tissue engineering. A well-designed scaffold for bone tissue engineering should possess biocompatibility, biodegradability, mechanical strength, and osteoconductivity. For this purpose, mesoporous silica nanoparticles (MSNs) were synthesized and incorporated at different ratios (10, 20, and 30%) into alginate/chitosan (Alg/Chit)-based porous composite scaffolds fabricated through the freeze-drying method. The MSN incorporation significantly improved the mechanical strength of the scaffolds while showing a negligible decreasing effect on the porosity. All of the samples showed desirable swelling behaviors, which is beneficial for cell attachment and proliferation. The MSN-containing scaffolds indicated a decreased hydrolytic degradation in an MSN percentage-dependent manner. The fabricated scaffolds did not depict cytotoxic characteristics. The Alg/Chit/MSN30 scaffolds not only showed noncytotoxic properties, but also increased the cell viability significantly compared to the control group. The biomineralization properties of the MSN-containing nanocomposite scaffolds were significantly higher than the Alg/Chit composite, suggesting the potential of these nanoparticles for bone tissue engineering applications. Taken together, it is concluded that the Alg/Chit/ MSN30 scaffolds are considerable substances for bone tissue regeneration, and MSN has a great tissue engineering potential in addition to its extensive biomedical applications.

## 1. Introduction

Bone defects caused by trauma, tumor, and congenital defects are increasing considerably. Accordingly, there is an increasing demand for bone tissue regeneration. Furthermore, small periodontal and peri-implant defects to large and critical lesions caused by trauma, tumor resection, and congenital malformations are all instances of oral and craniofacial bone defects that significantly impact both esthetics and functions in the mouth and face [1–3].

Scaffolds have an essential role in bone tissue engineering as a three-dimensional (3D) support for tissue formation by providing a proper microenvironment for cells to attach, proliferate, and osteogenic differentiation [4, 5]. Ideal scaffolds should mimic the structure and function of natural tissue; moreover, they should exhibit biocompatibility, controllable biodegradability, non-immunogenicity, and tailorable structural preferences as porosity, interconnected pore structure, and mechanical strength similar to natural bone [6–8]. In this regard, natural polymers play a pivotal role for enhancing the biocompatibility [9, 10]. For instance, scaffold based on natural polymers have attained great attention for bone tissue engineering owing to their desirable biocompatibility, biodegradability, and low immunogenicity. Chitosan is a naturally derived polymer and is deacetylated form of chitin. It shows superior properties, including availability, biocompatibility, biodegradability, and antimicrobial characteristics [11–15]. Alginate is a polysaccharide with advantageous features, such as biocompatibility, biodegradability and non-immunogenicity [16, 17].

Despite the above-mentioned advantages of the natural polymers, these polymers show poor mechanical characteristics, limiting their applications as a bone tissue engineering scaffold. Composite scaffolds have been introduced as an approach to address these limitations. Composite scaffolds consist of at least two materials with different characteristics to cover their drawbacks and achieve a scaffold with superior properties [18, 19]. Due to covalent interactions between the amine groups of chitosan with the carboxyl groups of alginate, it is suggested that chitosan and alginate are combined to create a composite scaffold with superior mechanical properties [20].

Furthermore, the addition of nanomaterials to polymeric scaffolds has demonstrated to improve bone regeneration capacity, mechanical properties, and swelling behavior [21]. For instance, mesoporous silica nanoparticles (MSNs) demonstrated unique merits, including large specific surface area, adjustable pore volume, controllable size, easy surface functionalization, and good biocompatibility, explaining the increasing biomedical application of MSNs. MSNs are extensively applicated in delivery systems, bioimaging, biosensing, and therapeutic agents. The employment of MSNs for tissue engineering applications is a relatively emerging subject that has sparked interest in the scientific community [22, 23]. These mentioned properties of MSNs besides their low cost and availability merit their application in the combination of tissue engineering with promising drug, gene, and protein delivery characteristics. Hence, fabricating biodegradable multifunctional scaffolds for promoted bone tissue engineering besides incorporating novel carrier agents for potential simultaneous delivery applications is the novelty of this project.

In light of this, we fabricated a biocomposite scaffold comprising chitosan and alginate possessing promising physical and biological properties of the composite scaffolds for bone tissue engineering applications. Besides, to improve the physical characteristics and osteogenic differentiation potential, we synthesized and incorporated MSN in the scaffolds. Subsequently, the physicochemical properties of the synthesized nanoparticles and the host scaffold were investigated. Following that, cellular assays, such as cytotoxicity, alkaline phosphatase activity, and calcium deposition evaluation were carried out to evaluate its capability in bone tissue regeneration.

## 2. Materials and methods

### *2.1. Materials*

The solvents and materials were purchased from Merck (Germany), Sigma-Aldrich (USA), and Abcam (USA) unless otherwise noted.

### *2.2. Preparation of MSNs*

A solution containing 0.400 gr cetyltrimethylammonium bromide (CTAB)/192 ml DI water/1.4 ml NaOH (2 M) was stirred at 50 °C for 30 min at 1000 rpm. Then, 2.8 ml mesitylene was added, and the temperature was set at 80 °C. An aluminum foil was wrapped around the container after the mesitylene addition to prevent the undesirable light effect. After 4 h, 2 ml tetraethoxysilane (TEOS) was added dropwise to the solution and continued to be stirred for 2 h. Afterward, a filter paper was placed in a Büchner funnel connected to a Büchner flask with a tube connected to a vacuum suction. The prepared solutions from the previous steps were poured separately on the filter paper, and the remaining mixture was incubated at 60 °C for 24 h and then at 120 °C for 24 h to remove the water. Then, the samples were calcined for 4 h at 550 °C. Finally, the powders were manually ground using a ball mill to obtain a homogenous MSN powder (Fig. 1A).

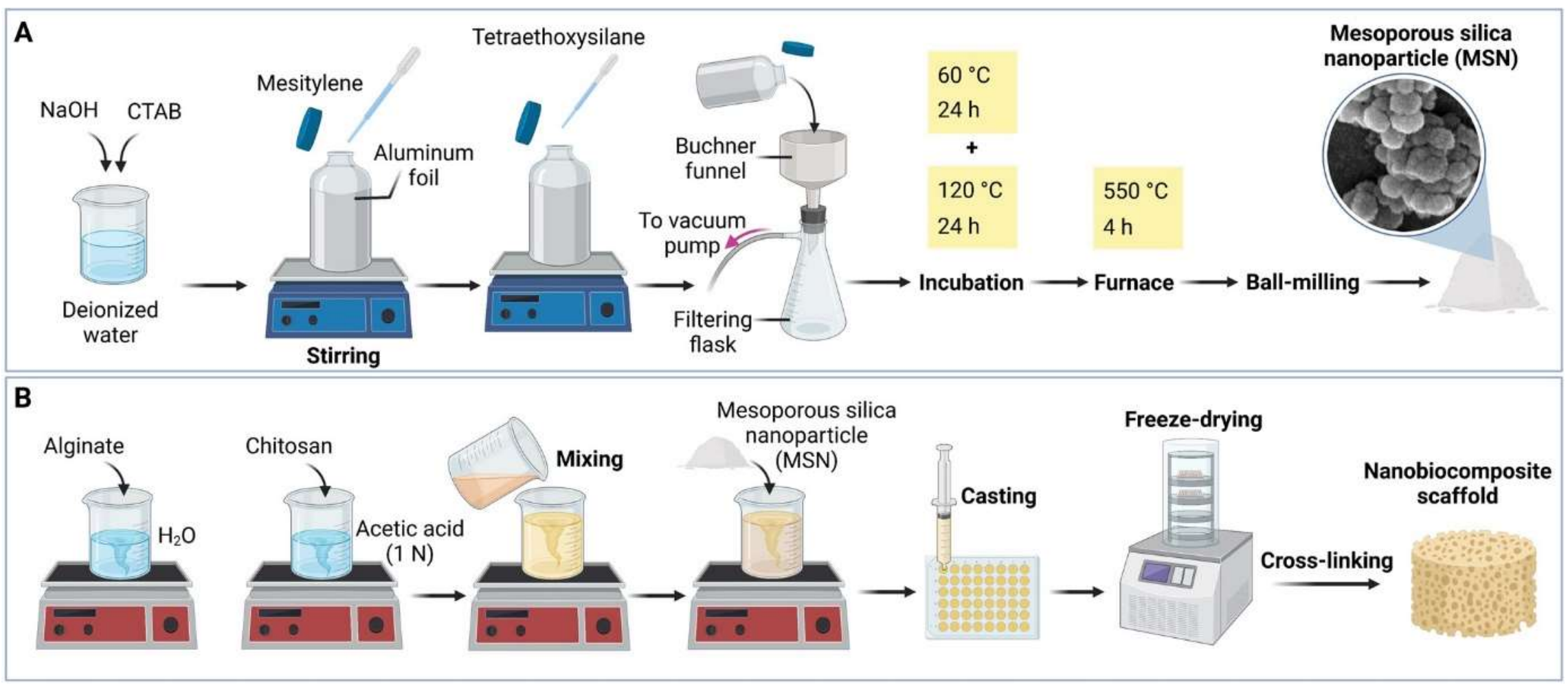

**Fig. 1.** Schematic illustration of the (**A**) synthesizing steps of the MSNs and (**B**) fabrication process of the Alg/Chit-based nanocomposite scaffolds.

### *2.3. Nanoparticle characterization*

#### FTIR

The functional groups of the synthesized MSNs were identified using Fourier-transform-infrared spectroscopy (FTIR, Bruker, Germany). Briefly, 1 mg of each sample was mixed with 300 mg KBr as a standard reference, pelleted, and analyzed at 2.60 Hz with a resolution of 4 $cm^{-1}$ in the wavenumber range of 4000–400 $cm^{-1}$.

#### XRD

The X-ray diffraction (XRD) method was used to identify different phases present in the synthesized MSNs with a Philips X-ray diffractometer using the Cu-Kα radiation (λ = 1.54056 Å) with the step size of 0.05°, step time of 2 s, and scanning range of 5–70 degrees.

#### FESEM and EDS

The size, morphology, and elemental analysis of the nanoparticles were examined by field-emission scanning electron microscopy (FESEM, MIRA3, Tescan, Brno, Czech Republic) equipped with energy dispersive X-ray spectrometry (EDS).

#### Dynamic light scattering (DLS)

The synthesized MSNs were well dispersed in DI water using sonication. Hydrodynamic particle sizes and size distributions were acquired using a dynamic light scattering instrument (DLS, Malvern Panalytical, Worcestershire, UK).

### *2.4. Preparation of porous chitosan–alginate and MSN scaffolds*

An alginate solution was made by dissolving alginic acid sodium salt powder in deionized $H_2O$ to prepare a 5% (wt./v) solution. Chitosan (deacetylation degree < 80%, low MW) was dissolved in acetic acid (1 N) to prepare 5% (wt./v) chitosan solution. Then, 50% alginate and 50% chitosan solutions were mixed under vigorous stirring at room temperature. For the fabrication of composite scaffolds, an appropriate amount of the MSNs was added to the chitosan-alginate solution to obtain samples containing 10–30% MSNs (Alg/Chit/MSN 10%, 20% & 30%). To ensure homogenous dispersion and prevent aggregation of MSNs in the polymer solution, the ball-milled MSN powders were sonicated in deionized water until obtaining a homogenous mixture. The solution underwent 6 h of stirring to reach a homogenous solution. Then, the homogenous solutions were pured in 48 well plate, stored at 4 °C for 2 h, and immediately moved to − 30 °C for 24 h. Before freeze-drying, the samples were moved to − 80 °C for one hour and then lyophilized in a freezer dryer (Alpha 1–2 LD plus, Christ,

Germany) at − 50 °C and 0.040 bar for 48 h. Then, the fabricated scaffolds were sliced in desired thicknesses. The scaffolds were cross-linked by immersing in a $CaCl_2$ solution (1% wt./ml) for 15 min, rinsed with DI $H_2O$, and moved to − 30 °C after 1 h of incubation at 4 °C. Finally, the scaffolds were lyophilized in the freeze dryer (Alpha 1–2 LD plus, Christ, Germany) at − 50 °C and 0.040 bar for 48 h (Fig. 1B).

### *2.5. Characterization of the fabricated nanocomposite scaffolds*

#### SEM and EDS

A thin layer of gold was sputter-coated (SCD 004, Balzers, Germany) on the samples before microscopic analyses to ensure an effective conductivity on the surface avoiding surface charging. Scanning electron microscopy (SEM, TESCAN VEGA3 XMU, Tescan, Brno, Czech Republic) equipped with energy dispersive X-ray spectrometry (EDS) was used to investigate the morphological and semi-quantitative elemental properties of the nanocomposite scaffolds at a 20 kV accelerating voltage.

#### ATR-FTIR

The functional groups of the fabricated scaffolds were identified using Attenuated Total Reflectance-Fourier Transform Infrared spectroscopy (ATR-FTIR, Bruker, Germany). Briefly, the dried scaffold samples were cut into 10 mm diameter and 1 mm thickness. They were employed for analysis with a resolution of 4 $cm^{-1}$ in the wavenumber range of 4000–600 $cm^{-1}$.

## Porosity and pore size measurement

By analyzing electron microscope images using the Image J software, the porosity and pore distribution of the scaffolds were determined. To determine porosity, the surface area of pores was divided by the total image area. This measurement was carried out on a series of SEM images, and the mean results are expressed. In addition, the average pore size and pore size distribution were reported after measuring the size of the pores randomly in different SEM pictures (at least 100 pores).

## Mechanical testing

The compressive mechanical characteristics of the fabricated scaffolds were evaluated using monotonic uniaxial unconfined compression, regarding a procedure mentioned elsewhere [24]. At least three scaffolds of each composition were prepared in a cylindrical shape with 10 mm height and 10 mm diameter. The samples were immersed in the phosphate buffer saline (PBS) for one hour before the test to preserve the physiological conditions of the tissue. The tests were performed at a 1 mm/min rate at room temperature using a universal mechanical tester (SANTAM, STM-20, Iran). The elastic modulus of each sample was determined from the slope of engineered stress–strain curves.

## Swelling behavior

The swelling behavior of the fabricated scaffolds was measured according to a protocol described elsewhere [25]. The initial dry weight of each scaffold with 10 mm diameter and 5 mm thickness was first measured, immersed in the phosphate-buffered saline solution (PBS), and incubated in vitro at 37 °C. Afterward, the wet weight of the samples at predetermined time intervals (5, 10, 20, 30, 40, 50, 60, 120, 240, 360, and 1440 min) was measured after removing the excess water from the surface of the scaffolds. The scaffold swelling ratio was calculated using the following equation
(Swelling ratio = $W_w$—$W_d$/$W_d$), where $W_w$ is the weight of the wet sample, and $W_d$ corresponds to the initial dry weight of the same sample. Data from three samples were obtained and were expressed as mean ± SD.

**In vitro *degradation assay***

The in vitro degradation assay was performed according to a previous report [25]. The samples were immersed in PBS for 2 h to attain swelling equilibrium before weighing for the in vitro hydrolytic degradation tests. This record was considered as the initial weight, and the scaffolds were immersed in PBS (pH 7.4) and incubated at 37 °C for 21 days. Two thirds of the PBS solution was changed with a fresh solution every three days. At predetermined time points (3, 7, 14, and 21 days), the samples were withdrawn from PBS, and the weight of each sample was measured. The scaffold degradation percentage was determined by the following equation: Degradation (%) = ($W_t$ – $W_i$ / $W_i$) × 100, where $W_i$ corresponds to the initial weight of the sample and $W_t$ is the weight of the sample at various time periods. Data from three scaffolds were obtained and expressed as mean ± SD.

**Cell toxicity analysis**

The biocompatibility of the synthesized MSNs and fabricated scaffolds was assessed by in vitro cytotoxicity tests using the 3-(4,5-dimethylthiazol-2-yl)-2, 5-diphenyl tetrazolium bromide (MTT) assay. Rat bone marrow mesenchymal stem cells (BMSCs) were cultured in the DMEM medium supplemented with 10% fetal bovine serum and 1% penicillin/streptomycin antibiotic and incubated at 37 °C with 5% $CO_2$. The MSN powder was well dispersed in the culture media at the concentrations of 250, 500, 750, 1000, and 1500 µg/ml, incubated for one day at 37 °C, and sterilized by filtration. BMSCs were seeded on 96-well culture plates at a density of $3 \times 10^3$ cells per well. The culture medium was replenished with the particle suspensions after cellular attachment, and this period was set as the beginning of culture time. The suspensions were refreshed every other day. After predetermined periods (1, 3, and 5 days), the MTT solution (0.5 mg/ml) was added to each well and incubated at 37 °C for 4 h. Then, the medium of each well was removed, and 100 µL Dimethyl sulfoxide (DMSO) was added to each well to dissolve insoluble formazan crystals. The plates were gently shaken for 15 min in the absence of light, and formazan product absorbance was measured by a microplate spectrophotometer (Tecan's Sunrise, Austria) at a wavelength of 570 nm.

The scaffolds were sectioned in a 1 mm thickness and 6 mm diameter with an approximate weight of 2.4 mg. The samples were sterilized using 70% ethanol followed by 2 h of UV radiation and then were washed twice with PBS and culture media. The prepared scaffolds were placed in a 48-well plate, and the MSCs were seeded at a density of $5 \times 1\ 0^4$ cells per scaffold on the top of each scaffold. The culture media was replaced every other day. After predetermined periods (3, 5, and 7 days), the MTT solution (500 µl/ml) was added to each well and incubated at 37 °C for 4 h. After that, the medium of each well was removed, and 150 µL Dimethyl sulfoxide (DMSO) was added to the each well to dissolve insoluble formazan crystals. The plates were gently shaken for 15 min in the absence of light and then moved to a 96-well plate. Finally, the formazan product absorbance was measured by a microplate spectrophotometer (Tecan's Sunrise, Austria) at a wavelength of 570 nm.

## Osteogenic differentiation

The scaffolds were sectioned in a thickness of 1 mm and diameter of 10 mm and were sterilized using a UV light for 1 h. Then, each scaffold was placed in a 24-well culture plate. First, 300 µL of complete medium (DMEM containing 10% FBS and 1% Penicillin–Streptomycin) was dropped on top of the scaffolds, and the plate was kept at 37 °C, 5% $CO_2$ incubator for an hour before cell seeding. Afterward, MSCs were dropped onto the scaffolds with an average density of $10^5$ cells per well. The plate was placed back in the incubator (37 °C, 5% C $O_2$) for 10 min, and the volume of the medium in the wells was raised to 1000 µL with the complete medium. The complete medium was interchanged one day after seeding by osteogenic media (BioIdea, Iran). After that, for the next 21 days, half of the osteogenic media in the wells was refreshed every day. Measuring the calcium content and alkaline phosphatase level was used to investigate the differentiation process at different time intervals of 3, 7, 14, and 21 days.

## Alizarin red staining

The alizarin red staining assay was used to determine the calcium content composed onto the scaffolds during the osteogenic differentiation. To summarize, the culture media was discarded, and the scaffolds were rinsed twice using PBS. To fix the scaffolds, they were put in 4% formaldehyde for 15 min and then rinsed using PBS. Afterward, 300 µL of 1.5% alizarin red (pH 4.2) was added to the scaffolds and incubated for 4–5 min at room temperature. Then, the scaffolds were rinsed several times with PBS (pH 4.5) to remove any remaining dye. Qualitative assessments were performed by taking images using an invert microscope (Motic, China). Subsequently, the scaffolds were incubated with an acetic acid solution (8%) for 20 min at room temperature under gentle shaking to solve deposited dyes. Then, the plates contents were moved to a 96-well plate, and the absorbance was measured at a wavelength of 550 nm. The optical absorption of various concentrations of alizarin red was measured at 550 nm wavelength to obtain a standard curve. The calcium content was assessed by the fact that an mole of alizarin red interacts with two mole calcium [26].

**Alkaline phosphatase assay**

ALP enzyme activity was measured through the p-nitrophenyl phosphate (pNPP) method using Pars Azmoon kit (Iran). To begin, the culture media was removed, and the scaffolds washed in PBS. The cell-containing scaffolds were lysed by NP40 buffer. The lysed contents were transferred to a microtube and centrifuged for 10 min at 3000 rpm. The microtubes supernatant were moved into a 96-well plate, the ALP substrate was added and reaction was halted after 30 min of incubation at 37 °C. Eventually, the plate absorbance was measured at 405 nm and the ALP enzyme activity was calculated as Unit/L.

### *2.6. Statistical analysis*

Data are reported as mean ± standard deviation (Mean ± SD) and were analyzed using the GraphPad Prisma 9 software (San Diego, CA, USA). Statistical data analyses were performed using one-way analysis of variance (ANOVA) followed by post hoc Tukey's tests. A *p*-value below 0.05 was considered statistically significant for all the tests.

## 3. Results and discussion

### *3.1. Characterization of synthesized MSN*

The Fourier-transform-infrared (FTIR) spectrum of the synthesized MSNs is shown in Fig. 2A. The FTIR spectroscopy indicates that the MSNs present peaks at 455 and 814 $cm^{-1}$ corresponding to Si–O bonds. The peak observed at 1088 $cm^{-1}$ also corresponds to the Si–O-Si vibrations of silanol groups. The peaks range of 3200–3400 $cm^{-1}$ is attributed to the O–H bond. In the XRD pattern of the MSNs (Fig. 2B), the broad diffraction peak detected at 2 theta = 23º with a hollow feature confirms an amorphous nature, which is related to the PDF card standard of silica-cristobalite phase ICDD #00-001-0424 [27]. FESEM images show that the synthesized MSNs have a monodispersed spherical-like shape with a diameter of approximately 100 nm (Fig. 2C). The FESEM results also indicate that the sample has a narrow particle size distribution. The EDS analysis depicts the presence of Si and O in the synthesized MSNs, which confirms that the synthesized particles are pure MSNs without impurities (Fig. 2D). The DLS measurements also showed an average size of about 270 nm with a narrow particle size distribution (Fig. 2E). The particle size observed by FESEM is smaller than the hydrodynamic diameter measured by DLS. Variations in DLS and FESEM results are commonly indicated in the literature since DLS provides a hydrodynamic diameter; thus, the measured particle size is expected to be larger than the exact particle size.

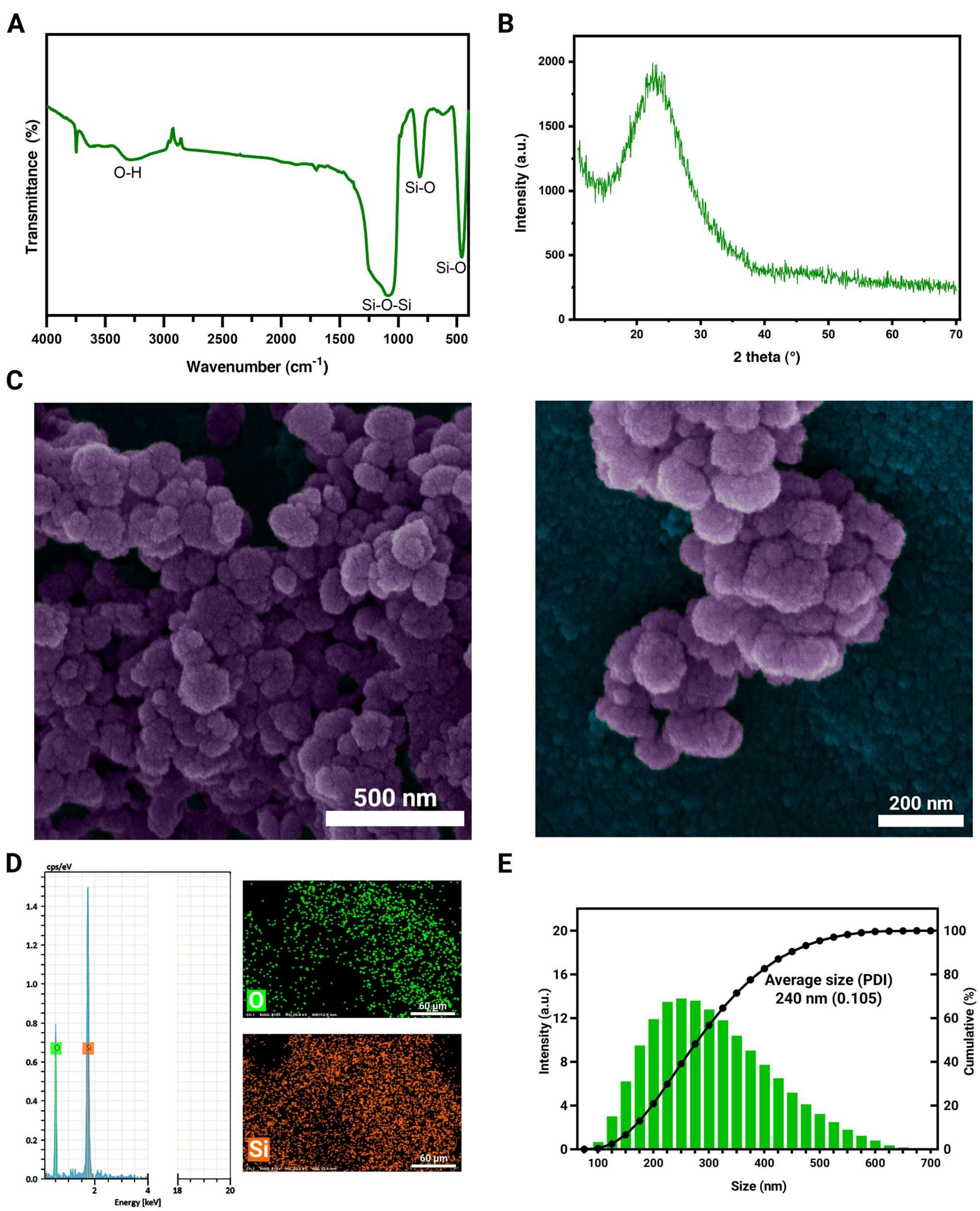

**Fig. 2.** MSN characterization results: (A) FT-IR spectrum of the synthesized MSNs, (B) XRD pattern of the MSNs, (C) representative FESEM micrographs of the MSNs, (D) EDS analysis of the MSNs, and (E) average hydrodynamic size and hydrodynamic size distribution of the synthesized MSNs.

Furthermore, the rare presence of large particles besides weak aggregates of the particles can cause the z-average particle size to increase [28].

### *3.2. Characterization of the fabricated composite scaffolds*

The SEM micrograph of the scaffolds reveals the porous structure of the four types of the scaffolds. The porous and rough surface structure of the scaffolds are ideal for stem cell adhesion and shaped cell growth. The EDS elemental mapping and profiles also confirm the incorporation and uniform distribution of the MSNs (the presence of Si element) within the scaffolds (Fig. 3).

The ATR-FTIR spectra of the non-crosslinked Alg/Chit, crosslinked Alg/Chit, Alg/Chit/MSN10, Alg/Chit/MSN20, and Alg/Chit/MSN30 are represented in Fig. 4A. The bands at 1424 and 1070 $cm^{-1}$ in the Alg/Chit scaffold are attributed to –COOH and C–O stretching bands, respectively. In the cross-linked Alg/Chit spectra, amide II bands were considerably intensified. The stretching vibrations of the C=O groups of the prepared scaffolds are assigned to 1609 $cm^{-1}$, 1617 $cm^{-1}$ and 1624 $cm^{-1}$, respectively. The peak observed at 1088 $cm^{-1}$ corresponds to the Si–O–Si vibrations of silanol groups, confirming the presence of the MSNs in the Alg/Chit scaffolds.

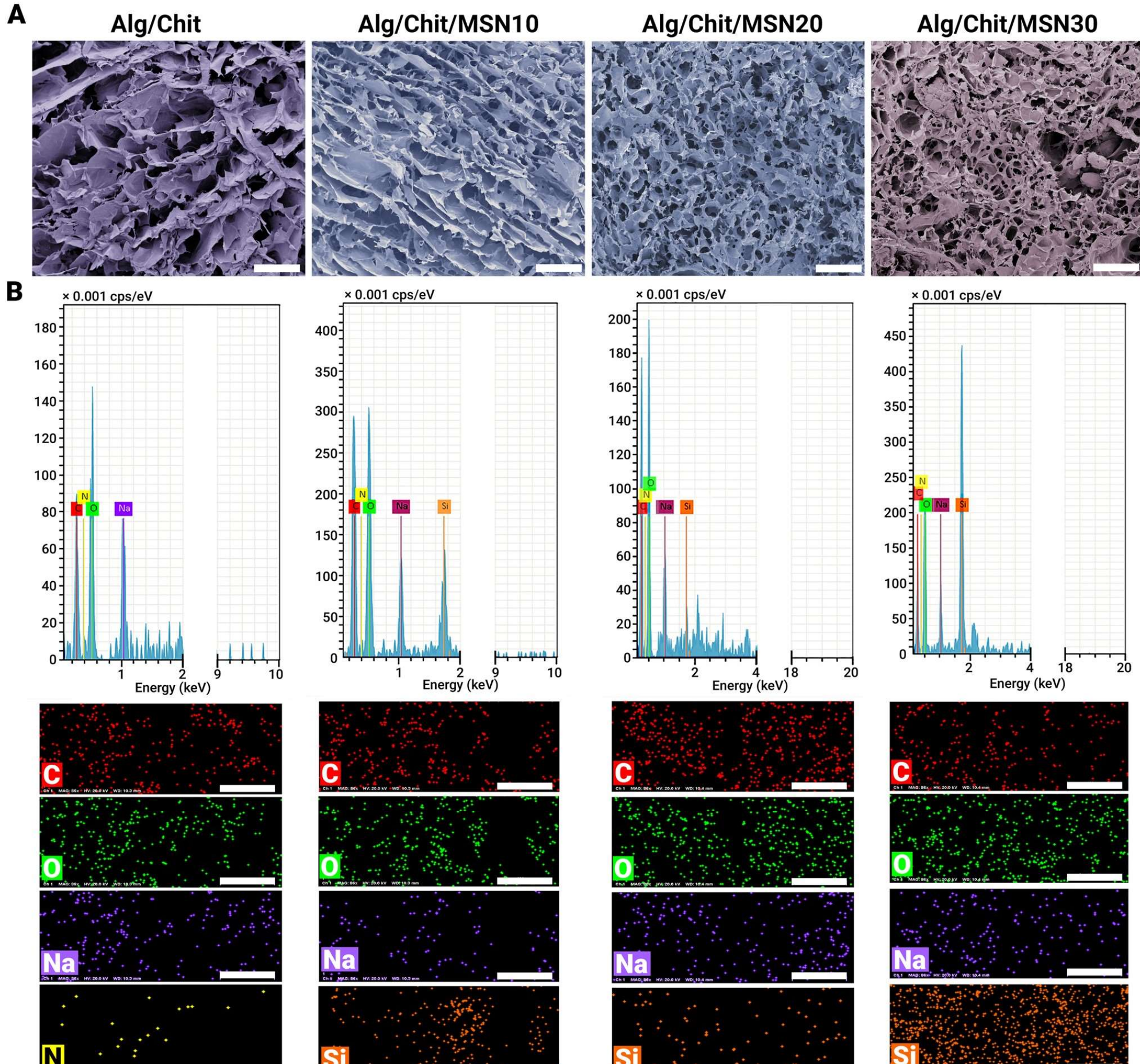

**Fig. 3.** (**A**) SEM micrographs of the fabricated scaffolds (white scale bars indicate 500 μm) and (**B**) corresponding EDS analyses of the fabricated scaffolds (white scale bars indicate 400 μm)

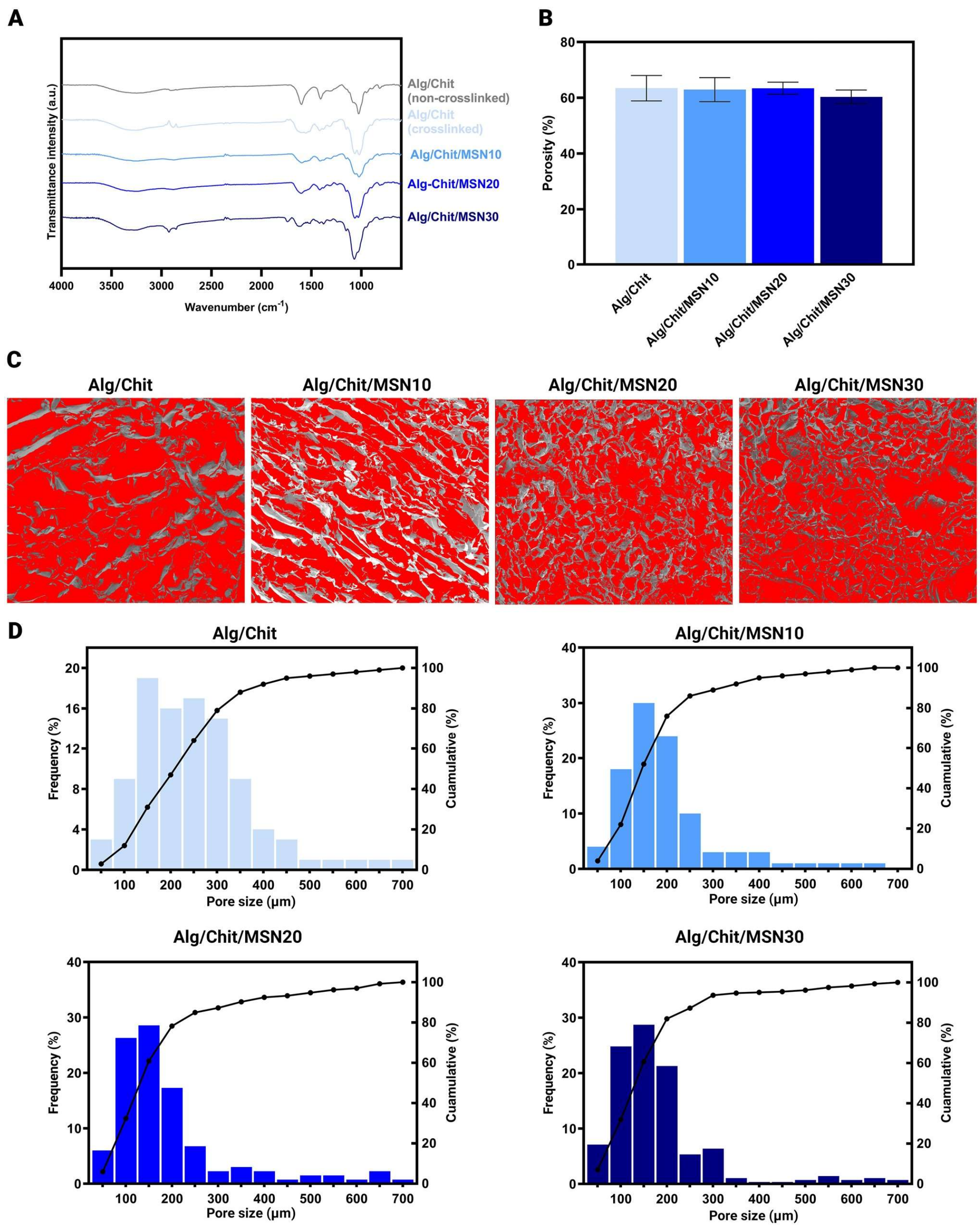

**Fig. 4.** (**A**) ATR-FTIR spectra of the non-crosslinked Alg/Chit, crosslinked Alg/Chit, Alg/Chit/MSN10, Alg/Chit/MSN20, and Alg/Chit/MSN30 scaffolds, (**B**) graph and (**C**) corresponding images of the fabricated scaffolds porosity, and (**D**) pore size distribution of the scaffolds

The porosity of the fabricated Alg/Chit, Alg/Chit/ MSN10, Alg/Chit/MSN20, and Alg/Chit/MSN30 scaffolds was measured to be 63.4 (± 4.5) %, 63.4 (± 2.1) %, 62.9 (± 4.3) %, and 60.3 (± 2.5) %, respectively (Fig. 4B,C). The pore size of the fabricated scaffolds ranges from less than 50 µm to 700 µm. The average pore size of the Alg/ Chit, Alg/Chit/MSN10, Alg/Chit/MSN20, and Alg/Chit/ MSN30 scaffolds are 221, 160, 124, and 119 µm, respectively (Fig. 4B).

## Mechanical testing

The mechanical stability of the scaffolds was determined by measuring mechanical strength against vertical dimensional change and calculating elastic module. We measured the mechanical stability after soaking the scaffolds in the PBS solution to resemble the normal body conditions (Fig. 5A,B). Mechanical strength is vital for in vivo applications and is between 0.3 and 1.3 MPa in our fabricated scaffolds, so they are more suitable for engineering cancellous bone rather than cortical bone. However, soaking in PBS significantly decreases the mechanical stability of scaffolds [29]. Indeed, while bulk mechanical parameters for osteogenic scaffolds intended to function as biodegradable temporary structures are different from those for load-bearing dense permanent implants, swelled scaffolds should exhibit sufficient stiffness to enable surgical handling and stable graft placement [25].

The minimum mechanical strength and elastic module belong to the Alg/Chit scaffold. In our experiments, the mechanical stability is significantly increased with the MSN content in the polymer structure. In this regard, it is reported that adding MSNs into polymer scaffolds will increase the strength of composites since MSNs lead to an efficient mechanical stress transfer mechanism. Also, it has been revealed that the porous nature of MSNs leads to a mechanical interlocking among the nanoparticles and between the particles and polymer chains, giving the polymer an opportunity to form stronger bonds [30].

**Swelling behavior**

Water absorption capacity is critical for tissue engineering scaffolds, as the procedure is dependent on biological solution absorption, nutrient penetration, and metabolic wasted transport within the biomaterials [31]. The above-mentioned characteristics can be maintained using a high swelling ratio in the fabricated scaffolds.

In terms of water uptake by the dry scaffolds, all the samples demonstrate a high swelling capacity (Fig. 5C). After soaking for the extended periods, the increased swelling ratio indicates an increase in the water absorption amounts; however, the increasing water absorption rate is decreased, and the scaffold reached a swelling equilibrium. The Alg/ Chit scaffold shows the highest swelling rate compared to the nanocomposite scaffolds. Even though the incorporation of the MSNs does not alter the swelling kinetics (all the samples are swelled after 1 h), the swelling ratio is decreased with increasing the reinforcement levels. The decrease in the swelling ratio with increasing the MSN concentration can be attributed to the porosity changes of the scaffolds and the relative reduction of the polymer concentration in the scaffolds. Because the biopolymer swelling capacity is generally related to the interaction of polymer components with water molecules. Furthermore, these results are in accordance with the presented results of recent studies on the swelling ratio of Alg/Chit composite and nanocomposite scaffolds [32, 33].

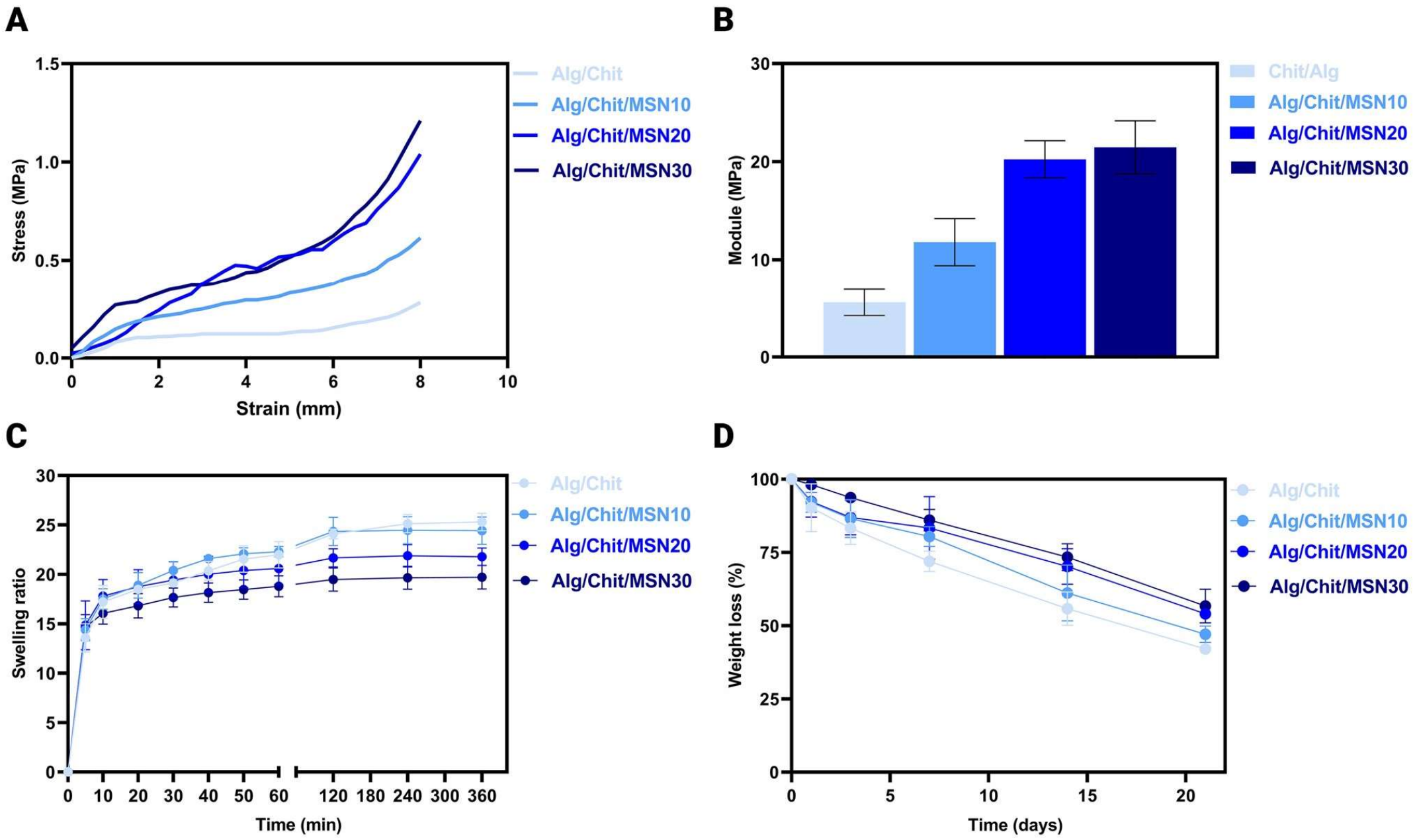

**Fig. 5.** Mechanical behavior of the fabricated scaffolds based on: (**A**) stress–strain curve and (**B**) elastic modulus. (**C**) Swelling ratio of the freeze-dried scaffolds and (**D**) In vitro hydrolytic degradation of the fabricated scaffolds

### In vitro *degradation assay*

The degradation behavior of scaffolds is critical for tissue regeneration and is regarded as one of the most desired characteristics for a scaffold developed for bone tissue regeneration. Scaffold degradation should occur at a nearly identical rate to that of tissue repair to support regenerative tissue ingrowth. The analysis of the scaffold stability under hydrolytic conditions for 21 days reveals that approximately 50% of the scaffolds were degraded after 21 days, and the Alg/ Chit and Alg/Chit/MSN30 scaffolds showed the highest and lowest degradation percentages, respectively (Fig. 5D). The results also indicated that the combination of the MSNs with the Alg/Chit scaffolds leads to a decrease in biodegradability, so that this decline was commensurate with the increase in the MSN percentage. This can be related to the low degradation rate of the MSNs [34]. Still, the in vitro recapitulation of the surrounding environment following the implantation of the scaffold is exceedingly complex due to the existence of multiple types of proteases that regulate extracellular matrix remodeling. Hence, the addition of MSNs could be beneficial for bone tissue engineering applications since it can prevent its rapid degradation of implants [35].

***Cell proliferation and viability* (*MTT*)**

The clinical translation of MSNs is conceivable only if they do not exhibit cytotoxic activity toward normal cell lines. The viability of cultured BMSCs on the synthesized MSNs was assessed by the MTT assay after 1, 3, and 5 days (Fig. 6A). On the 1st day, the MSNs with 1.5 mg/ml concentration show significantly decreased viability. However, all of the MSN concentrations do not induce significant cytotoxicity and have a similar cellular viability compared with the control group on 3rd and 5th days. Particle size and concentration play a key role in MSN cytotoxicity. It has been reported that MSNs ranging from 30 to 300 nm have no significant cytotoxicity toward HeLa cells [36]. Still, there is controversy regarding the results of MSN cytotoxicity, which can be related to variations in the MSN synthesis method and cytotoxicity measurement assays [37, 38].

It is essential for composites to remain inert in the human body and be compatible with bone formation when implanted. In addition, scaffolds should be biocompatible in addition to their physicochemical characteristics. The results indicated that all of the scaffolds show no toxicity compared with the two-dimensional cultured cells as the control group after 3, 5, and 7 days (Fig. 6B). Furthermore, the Alg/Chit/ MSN30 shows significantly higher cell viability compared with the control group after 3 and 5 days. The composition of the scaffolds largely influences biocompatibility characteristics. FDA has approved the use of chitosan and alginate as biomaterials; In fact, the biocompatibility of these constituent polymers is reflected in the cytotoxicity findings [39].

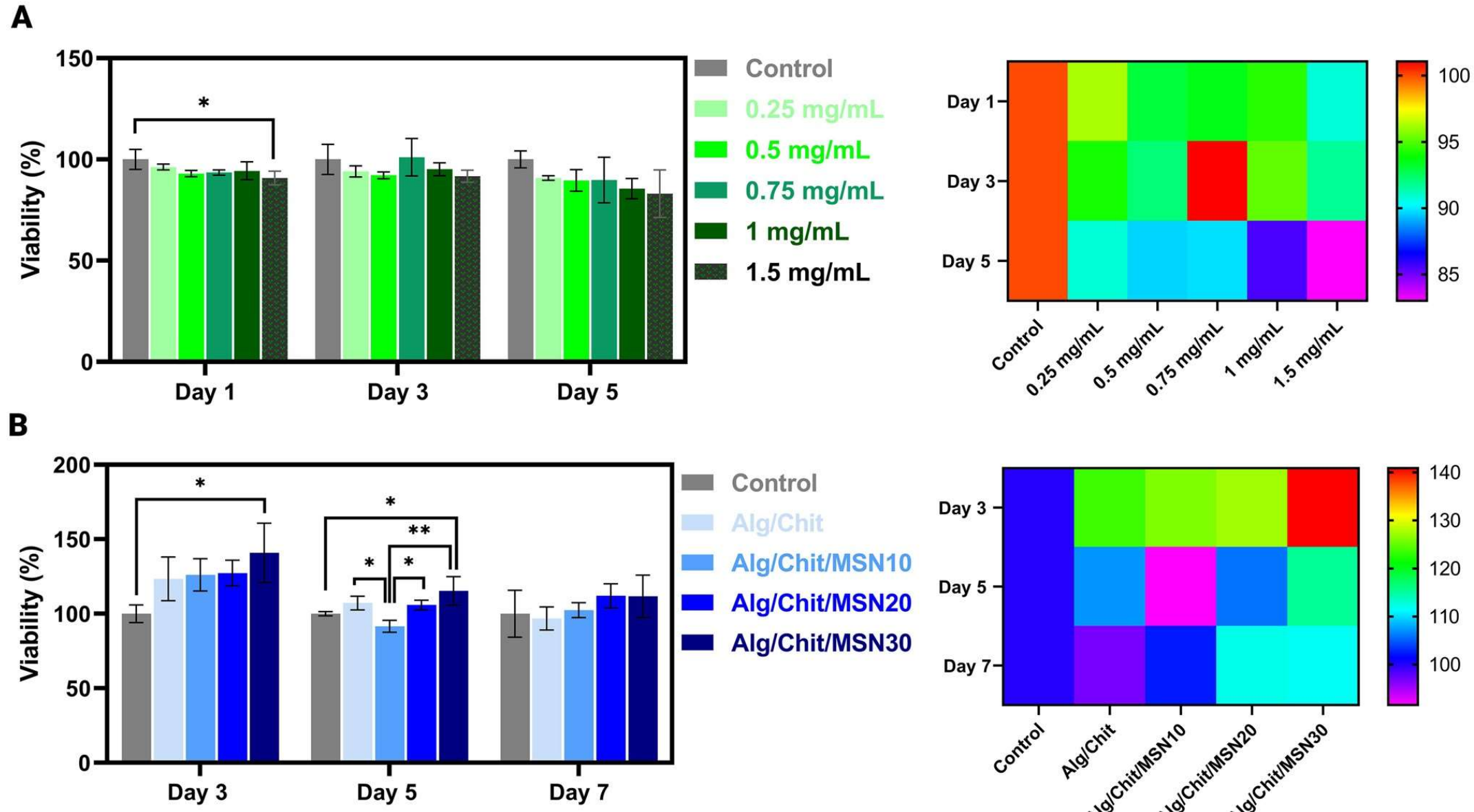

**Fig. 6.** (**A**) Cell viability of BMSCs cultured at 0.25, 0.5, 0.75, 1, and 1.5 mg/ml MSNs for 1, 3, and 5 days with the corresponding heatmap of the cell viability (*$p < 0.05$) and (**B**) cell viability of the two-dimensional cultured cells (control), Alg/Chit, Alg/Chit/MSN10, Alg/Chit/MSN20, and Alg/Chit/MSN30 scaffolds for 3, 5, 7 days with the corresponding heatmap of the cell viability (*$p < 0.05$, **$p < 0.01$)

## Osteogenic differentiation

The differentiation efficiency of the MSCs into mature osteocytes on the different scaffolds was examined using the alizarin red and alkaline phosphatase assays. The mineralization extent is higher on Alg/Chit/MSN20 than the other scaffolds, followed by Alg/Chit/MSN30 at 7, 14, and 21 days (Fig. 7A). Also, the content of calcium deposited on the Alg/Chit/MSN20 scaffolds is significantly higher than the Alg/Chit/MSN10 and Alg/Chit/MSN30 scaffolds at 7 and 14 days. However, there is no significant difference between the calcium deposition of the Alg/Chit/MSN20 and Alg/Chit/MSN30 scaffolds at 21 days. Based on previous studies, silica nanoparticles improve the osteogenic nature of scaffolds [40, 41] mainly through releasing biologically active ions, such as Si [42], inducing collagen production, enhancing ALP activity, and upregulation of bone-related genes, such as OPN RUNX2 and OCN [43]. Released Si ions can stimulate osteogenesis pathways, including TGFβ1[43], Wnt [44], ERK, and SHH signaling pathways [45]. Yang et al. [46] showed that silica nanoparticles significantly enhance the ALP activity and mineralization rate of human MSCs. Also, Ha et al. [47] reported that silica nanoparticles could stimulate the osteogenic differentiation of mouse MC3T3-E1 cells by increasing ALP activity, enhancing nodules formation, and upregulation of osteocyte-related genes. In the present study, the higher osteogenic efficiency of the Alg/Chit/MSN20 scaffold than Alg/Chit/MSN30 in 7 and 14 days can originate from a slight reduction of the cell proliferation rate on the Alg/Chit/MSN30 scaffold compared to the Alg/Chit/MSN20 scaffold in a time-dependent manner. However, the ALP activity in the Alg/Chit/MSN30 scaffolds is relatively higher than the others at all times. Since the ALP activity was normalized by the cell number, this test was not affected by the cell proliferation rate. A reduction is observed in the ALP activity of all the samples on 21 days. In this regard, previous studies reported that ALP activity is increased in osteoblasts and declined in mature osteocytes [48].

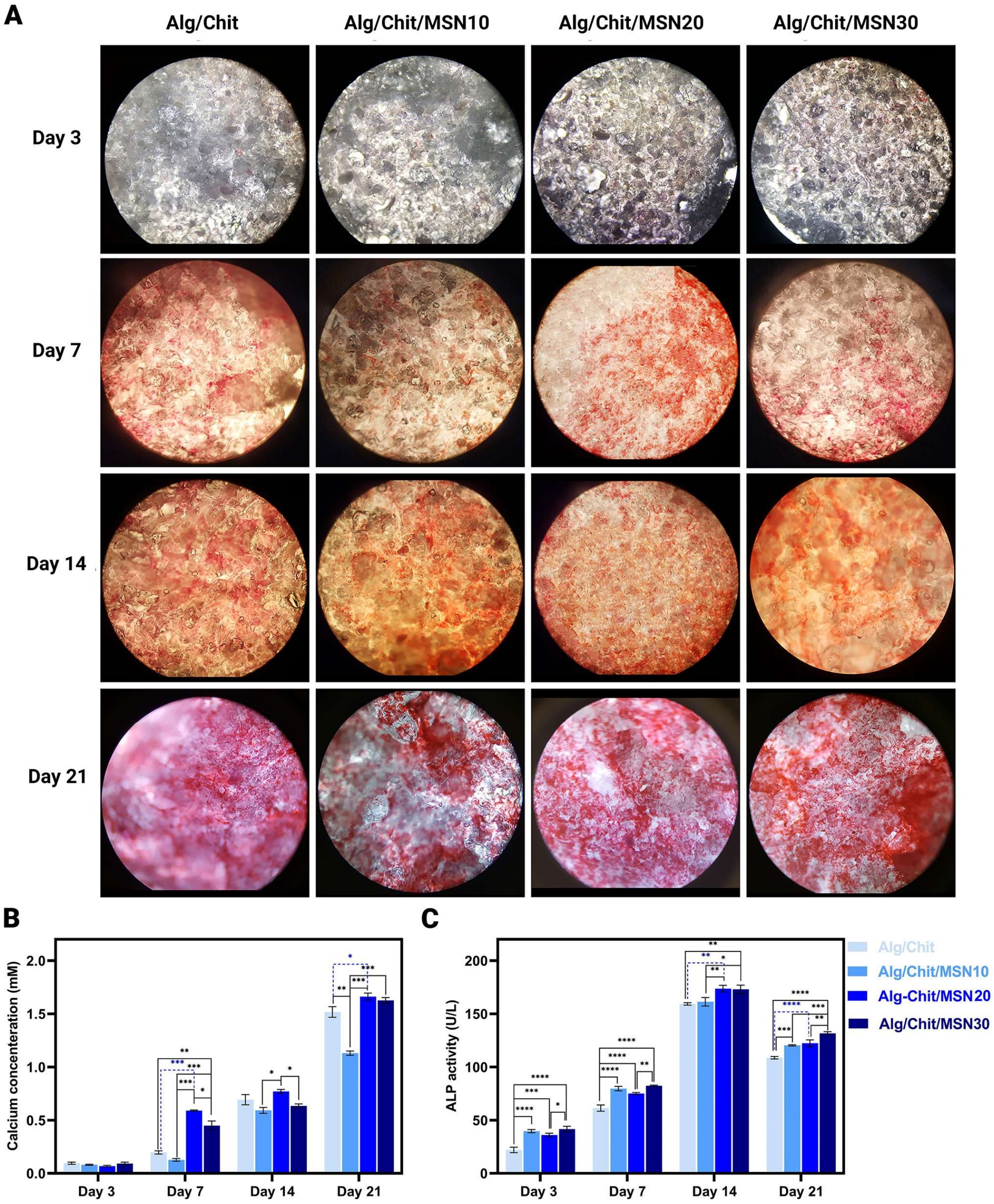

**Fig. 7.** Osteogenic potential of BMSCs cultured on the fabricated scaffolds: (A) Alizarin red staining images and (B) relative calcium content quantification after 3, 7, 14, and 21 days of cell culture (C) ALP activity secreted by BMSCs after 3, 7, 14, and 21 days of culture. (*$p < 0.05$, **$p<0.01$, ***$p< 0.001$, ****$p< 0.0001$)

## 4. Conclusions

In the present study, we successfully designed a nanocomposite scaffold via integrating MSNs into a Alg/Chit-based polymeric network. MSNs were synthesized and then employed to fabricate the nanocomposites with the varying concentrations of the MSNs (10, 20, and 30%). All of the fabricated scaffolds showed a proper swelling profile. The MSN-reinforced systems exhibited significantly increased mechanical properties with a slight decrease in porosity. Besides, the MSN incorporation into the polymeric structure prolonged the hydrolytic degradation that can depict the positive effect of these nanoparticles in the inhibition of rapid degradation upon implantation. The fabricated scaffolds showed no toxicity toward BMSCs and a slight decrease in the cell viability compared with two-dimensional cultured cells. In addition, Alg/Chit/MSN30 demonstrated significantly higher cell viability compared to the control group on 3rd and 5th day. Mineral deposition and alkaline phosphatase expression measurements indicated the positive effect of the MSN incorporation on osteogenic differentiation. Overall, the porous nanocomposite Alg/Chit/MSN scaffolds developed with desired characteristics could be a proper implant for craniofacial bone defects regeneration.

**Acknowledgements**

The study was funded by Vice-chancellor for Research and Technology, Hamadan University of Medical Sciences (Grant Number: 9804042537).

**This is the accepted manuscript (postprint) of the following article:**